\documentstyle[preprint,aps]{revtex}
\begin{document}
\preprint{gr-qc/9604021 }
\draft
\title{ Conformally dressed black hole in 2 + 1 dimensions }
\addtocounter{footnote}{1}
\author {
Cristi\'an Mart\'{\i}nez\thanks{Electronic address:
martinez@cecs.cl}}
\address{
Departamento de F\'{\i}sica, Facultad de Ciencias, Universidad
de Chile\\ Casilla 653, Santiago, Chile\\ and Centro de Estudios
Cient\'{\i}ficos de Santiago, Casilla 16443, Santiago 9, Chile.}
\author {
Jorge Zanelli\thanks{Electronic address: jz@cecs.cl}}
\address{
Centro de Estudios Cient\'{\i}ficos de Santiago, Casilla 16443,
Santiago 9, Chile \\ and Departamento de F\'{\i}sica, Facultad
de Ciencia, Universidad de Santiago de Chile,\\ Casilla 307,
Santiago, Chile.}
\date{ \today }
\maketitle
\begin{abstract}
A three dimensional black hole solution of Einstein equations
with negative cosmological constant coupled to a conformal
scalar field is given. The solution is static, circularly
symmetric, asymptotically anti-de Sitter and nonperturbative in
the conformal field. The curvature tensor is singular at the
origin while the scalar field is regular everywhere. The
condition that the Euclidean geometry be regular at the horizon
fixes the temperature to be $T=\frac{9\, r_+}{16\pi l^2}$. Using
the Hamiltonian formulation including boundary terms of the
Euclidean action, the entropy is found to be $\frac{2}{3}$ of
the standard value ($\frac{1}{4} A$), and in agreement with the
first law of thermodynamics.
\end{abstract}
\section{Introduction}  \label{introd}

In the last ten years, three dimensional gravity has become a
popular laboratory to understand the fundamentals of classical
and quantum gravity \cite{Carlip1}. Thus, the discovery of a
black hole solution in 2+1 dimensions \cite{BTZ} has further
contributed to the interest in three dimensional gravity. A
complete review about this black hole can be found in
\cite{Carlip2}. 

Several generalizations of this solution have been constructed.
For instance, minimally and non-minimally coupled dilaton field
with various black holes (charged and uncharged, spinning and
non-spinning) \cite{ChanMann1,ChanMann2,KleberLemosSa,Chan}, are
known. For other interesting extensions see \cite{Carlip2} and
references therein.

The purpose of this article is to report on an exact black hole
solution conformally coupled to a massless scalar field in 2+1
dimensions. The solution is static, circularly symmetric and
asymptotically anti-de Sitter and it possesses a curvature
singularity at the origin. The scalar field is regular
everywhere, has a fixed form and cannot be obtained as a
perturbation around a matter-free massive black hole. The system
can be shown to have a well-defined thermodynamic behaviour.

Here we consider gravity with cosmological constant conformally
coupled to a massless scalar field in $D$ dimensions. The action
is
\begin{eqnarray} 
I&= &I_G + I_C \, ,\, \mbox{with} \label{Daction}\\ I_G &=&
\frac{1}{2\kappa}\int d^D x \sqrt{-g} [R+2 l^{-2}]\,,\, 
\mbox{and}\\
I_C &=& -\frac{1}{2} \int d^D x \sqrt{-g} \left[ g^{\mu
\nu}\nabla_{\mu}\Psi\nabla_{\nu}\Psi -\xi_D \, R \,\Psi^2 \right]\,,
\end{eqnarray}
where $R$ is the scalar curvature and
$\xi_D=\frac{1}{4}(D-2)/(D-1)$.  The value of $\xi_D$ is chosen
so that $I_C$ be invariant under conformal transformations
\begin{equation} \label{conf}
g_{\mu \nu} \rightarrow \Omega^2(x) g_{\mu \nu} \qquad
\Psi \rightarrow \Omega^{1-\frac{D}{2}}(x)\Psi \,.  
\end{equation}

This coupling, including electromagnetism and without
cosmological constant in four dimensions, was previously
considered by Bronnikov, Melnikov and Bocharova \cite{BMB} and
Bekenstein
\cite{Bekenstein1,Bekenstein2} (see also \cite{Froyland}). The
uncharged BMBB black hole solution is static, spherically
symmetric and asymptotically flat (there is no cosmological
constant). The metric is the extreme Reissner-Nordstr\"om metric
solution and the scalar field is unbounded at horizon. In
\cite{Bekenstein2} it is
shown that this divergence is not physically troublesome.

Recently, the uniqueness of the BMBB black hole has been
established
\cite{XanthopoulosZannias} and it was shown to be the only static,
spherically symmetric, asymptotically flat black hole solution
of the Einstein-conformal field equations in four space-time
dimensions
\cite{XanthopoulosDialynas}.

Below we present a black hole solution for the system described
in Eq. (\ref{Daction}) in three dimensions. 

\section{Black hole solutions} \label{BHS}

In three dimensions, the action reads
\begin{equation} \label{action}
I= \int d^3 x \sqrt{-g} \left[ \frac{R+2 l^{-2}}{2\kappa}
-\frac{1}{2}g^{\mu \nu}\nabla_{\mu}\Psi\nabla_{\nu}\Psi
-\frac{1}{16}R
\,\Psi^2 \right]
\end{equation}
where $-l^{-2}$ is the cosmological constant and $\Psi$ is the
massless conformal scalar field.

The field equations are
\begin{equation} \label{Eeq}
G_{\mu \nu}-l^{-2}g_{\mu \nu}- \kappa \, T_{\mu \nu}=0
\end{equation}
and
\begin{equation} \label{Feq}
\Box \Psi-\frac{1}{8}R \Psi=0 \,, 
\end{equation}
where $\Box \equiv g^{\mu \nu}\nabla_{\mu}\nabla_{\nu}$ is
the Laplace -Beltrami operator in the metric $g_{\mu \nu}$ and
the matter stress tensor is
\begin{eqnarray} 
T_{\mu\nu}&=&\nabla_{\mu}\Psi\nabla_{\nu}\Psi-\frac{1}{2}g_{\mu\nu} 
 g^{\alpha \beta} \nabla_{\alpha}\Psi\nabla_{\beta}\Psi \nonumber \\
& &+\frac{1}{8}\left[g_{\mu \nu} \Box -
\nabla_{\mu}\nabla_{\nu} +G_{\mu \nu}\right]\Psi^2 \,. \label{Tuv}
\end{eqnarray}
It is straightforward to check that by virtue of Eq. (\ref{Feq})
the stress tensor is traceless. This in turn implies that the
geometry has constant scalar curvature, 
\begin{equation} \label{R}
R=-6l^{-2} .
\end{equation}

We look for static, circularly symmetric three dimensional
metrics whose expression in polar coordinates takes the form
\begin{equation} \label{ds}
ds^2= -N^{2}(r)F(r)dt^2+F^{-1}(r)dr^2+r^2 d\theta^2 \,,
\end{equation}
where $0 \leq r < \infty$ is the proper radial coordinate and $0
\leq \theta \leq 2\pi$.  The solution is easily obtained 
fixing the time scale so that $N(r)=1$. Working with the
advanced time coordinate $v=t+\int F^{-1}(r)dr$, the $r-r$
equation of (\ref{Eeq}) imposes the constraint 

\begin{equation} 
\label{psieq}
0= (\Psi^{'})^2-\frac{1}{8}(\Psi^2)^{''} \,,
\end{equation}
where prime denotes radial derivative. The above equation can be
written as $ 0= \Psi^4(\Psi^{-2})^{''}$ whose general solution
is
\begin{equation} \label{psi}
\Psi(r)=\frac{A}{\sqrt{r+B}} \qquad A,B \: \mbox{constants}.
\end{equation}
Comparing the curvature for the metric (\ref{ds}) with
(\ref{R}), one obtains directly
\begin{equation} \label{F}
F(r)=\frac{r^2}{l^2}-a-\frac{b}{r} \qquad a, b \:
\mbox{constants}.
\end{equation}
The ${}_v \,^v$ equation imposes the following relations among
the constants of integration
\begin{equation}
a=3B^2l^{-2} \qquad b=2B^3l^{-2} \qquad
A=\sqrt{\frac{8B}{\kappa}}
\qquad B \geq 0.
\end{equation}
Thus we obtain the black hole solution
\begin{equation} \label{Fp}
F(r)=\frac{1}{l^2}\left[r^2-3B^2- \frac{2B^3}{r}\right]=
\frac{(r+B)^2(r-2B)}{rl^2},
\end{equation}
together with the matter field configuration
\begin{equation} \label{psie}
\Psi(r)=\sqrt{\frac{8B}{\kappa (r+B)}} ,
\end{equation}
which can be explicitly checked to solve Eq. (\ref{Feq}). It is
easily shown in the advanced time coordinates that the surface
where F vanishes ($r=2B\equiv r_+$) is null \cite{footnote}.

The asymptotic behaviour of the metric is truly anti-de Sitter
(i.e., $g_{00} \sim r^2+ O(r^0)$, without terms linear in $r$).
Therefore, as shown in \cite{BrownHenneaux}, the asymptotic
symmetry group is the conformal one, which contains the anti-de
Sitter group as a subgroup.

The Riemann tensor is singular at the origin as can be shown by
evaluating the Kretschmann scalar
\begin{equation} \label{KR}
R^{\mu \nu \lambda \rho}R_{\mu \nu \lambda \rho}=\frac{
12(r^6+2B^6)}{l^4 r^6}.
\end{equation}
This is the only singularity and is hidden by the event horizon.

The massless conformal scalar field $\Psi$ is regular
everywhere. Although one might expect the scalar field to endow
the black hole with a hair, it should be noted that the solution
is characterized by only one constant which, as we will show
below, is related with the mass. Therefore, the presence of the
scalar field does not generate an independent additional charge
to the black hole, i.e., the scalar field produces no new hair.
Furthermore, the solution presented here does not differ in the
asymptotic region from a matter-free black hole.

\section{Thermodynamics} \label{Termo}

The Hamiltonian form of the action (\ref{action}) is given by
\begin{equation} \label{haction}
I = \int\left[ \pi^{ij}\dot{g}_{ij}+P\dot{\Psi} - N {\cal H}
-N^i{\cal H}_i \right] d^2xdt + B_{H}
\end{equation}
where $B_{H}$ is a surface term.

In order to study the thermodynamics of this system we consider
the minisuperspace of static, circularly symmetric geometries as
described by (\ref{ds}) and scalar fields that depend only on
the radial coordinate. The equations of motion obtained in this
way are the same as (\ref{Eeq},\ref{Feq}) after imposing the
above restrictions. Thus, reducing the Hamiltonian action
(\ref{haction}) to the minisuperspace gives
\begin{equation} \label{rhaction}
I = -2\pi(t_2-t_1)\int N(r){\cal H}(r) dr + B_{H}
\end{equation}
with
\begin{eqnarray} \label{rh}
{\cal H} &=&\frac{1}{2\kappa}[
F'(1-\zeta)-2Fr(\zeta''-\zeta^{-1}(\zeta')^2)\\
&&-(2F+F'r)\zeta'-2rl^{-2}],
\end{eqnarray}
\begin{equation} \label{zeta}
\zeta \equiv \frac{\kappa}{8} \Psi^2
\end{equation}

The partition function for a thermodynamical ensemble is
identified with the Euclidean path integral in the saddle point
approximation around the Euclidean continuation of the classical
solution \cite{GibbonsHawking}. In this approximation the
Euclidean action is related to the thermodynamic functions (in
units where $\hbar= k_B=1$ and $\kappa = 8\pi$) by 
\begin{equation} \label{appro}
I_E= \frac{ \mbox{free energy} }{ T}= \frac{M}{T}-S \,,
\end{equation}
where $T$, $M$, $S$ denote temperature, mass, entropy,
respectively and the Euclidean action $I_E$, is related to the
Lorentzian action by
\begin{equation} \label{e-l}
I_E= -i I \,, \qquad \tau= i t \,.
\end{equation}

The Euclidean continuation of the metric is
\begin{equation} \label{dse}
ds_E^2= N^2(r)F(r)d\tau^2+F(r)^{-1}dr^2+r^2 d\theta^2
\end{equation}
with $\tau_1 \leq \tau \leq \tau_2$ periodic, $r \ge r_+$, and
the scalar field unchanged.

The condition that the geometries allowed in the variation
should contain no conical singularities at the horizon implies
\begin{equation} \label{regular}
(\tau_2-\tau_1) F'|_{r=r_+}= 4\pi,
\end{equation}
which directly yields the temperature ($N(r)=1$)
\begin{eqnarray} \label{temp}
T\equiv \beta^{-1}&=&\frac{1}{\tau_2 - \tau_1} \\ &=& \frac{9
r_+}{16\pi l^2}.
\end{eqnarray}

We now turn to the evaluation of the Euclidean action at the
Euclidean solution. The classical solution is static and
satisfies the constraint ${\cal H}=0$ and therefore the action 
at the classical solution is given by a boundary term, $B_E$.
This boundary term must be such that the geometry (\ref{dse}) be
an true extremum among the class of metrics satisfying the right
boundary conditions \cite{ReggeTeitelboim,GaussBonnet}.

At infinity, we demand that the variations of the fields behave
as
\begin{eqnarray}
\delta N &=& 0, \\
\delta F & \rightarrow & - \delta \frac{3 r_+^2 }{4 l^{2}}, \\ 
\delta \zeta & \rightarrow & \frac{\delta r_+}{2 r}\,,  
\end{eqnarray}

At the horizon, we impose the regularity condition (\ref{regular})
\begin{equation} \label{regular2}
\beta F'\,|_{r=r_+}= 4\pi,
\end{equation}
and
\begin{equation} \label{hordef}
(\delta F )_{r_+} + F'\,|_{r=r_+} \delta r_+=0 \,,
\end{equation}
which is required by the definition of the horizon $F(r_+)=0$.
And, $ (\delta N)_{r_+}=0$.

The variation of the scalar field at the horizon is obtained by
varying it with respect to $r_+$, maintaining the functional
form of the classical solution, $\zeta= r_+ / (2r+r_+)$. Hence
\begin{equation} \label{dzetah}
\delta \zeta = \frac{2}{9 r_+} \delta r_+ \,.
\end{equation}

The variation of the Euclidean action is
\begin{eqnarray} 
\delta I_E & = & \frac{\beta}{8} [(1-\zeta-r\zeta')\delta F +
(F' r +4 Fr \zeta^{-1}\zeta') \delta \zeta \nonumber \\ & & -2Fr
\delta \zeta' ]_{r_+}^{\infty} + \delta B_{E} \nonumber \\
& & + \mbox{terms vanishing on shell}.
\end{eqnarray}
For convenience we write $B_E=B_E(\infty)+B_E(r_+)$. The
contribution from infinity is
\begin{equation} \label{Binf}
\delta B_E(\infty) = \beta \, \delta(\frac{3 r_+^2}{32l^2}) \,. 
\end{equation}
One can note here that the scalar field does not contribute to
surface term at infinity. This is yet another indication of the
non-existence the charges associated to the conformal scalar
field.

At the horizon, we have
\begin{equation} \label{Bhor}
\delta B_E(r_+) =  \beta \left[\frac{1}{9} \delta F + \frac{1}{36} F' 
\delta r_+ \right] \, ,
\end{equation}
which, in view of (\ref{regular2}) and (\ref{hordef}), can be
written as 
\begin{equation} \label{Bhor2}
\delta B_E(r_+) = -\frac{\pi}{3}\, \delta r_+ \,.
\end{equation}

Combining (\ref{Binf}) and (\ref{Bhor2}), the Euclidean action
is found to be
\begin{equation} \label{IE2}
I_E = \beta \frac{3 r_+^2}{32 l^{2}} -\frac{\pi}{3}\, r_+ + B_0
\,,
\end{equation}
where $B_0$ is an arbitrary constant independent of the fields
at the boundaries. Imposing that $I_E=0$ for $r_+\rightarrow 0$,
one finds that $B_0=0$. If we compare the above expression for
$I_E$ with (\ref{appro}) we learn that the energy and entropy
are
\begin{eqnarray}
M &=& \frac{3 r_+^2}{32 l^{2}} \\ S &=& \frac{\pi r_+}{3}  \,,
\end{eqnarray}
respectively. With these expressions, one can check that the
first law of thermodynamics
\begin{equation}
d M= T \, dS
\end{equation}
is satisfied.
\section{Concluding remarks} \label{resume}

The inclusion of the cosmological constant is absolutely
necessary for obtaining the black hole solution. Inspite of the fact
that the matter coupling in (\ref{Daction}) is of the same form as
that of the BMBB theory, the resulting black holes are entirely
different: The BMBB solution is asymptotically flat and is an extreme
Reisner-Nordstr\"om hole, whereas the solution introduced here is
asymptotically anti-de Sitter and non-extreme. Furtheremore, one can
readily see that the ansatz (\ref{ds})($N=1$) doesn't yield an
extension of the BMBB solution with cosmological constant in four
dimensions. 

The question of whether this solution represents a hairy black
hole depends on the definition of hair one uses. In a very broad
sense any matter field that can be sustained by a black hole
could be regarded as some kind of hair, as it is the case at
hand. However, in a more strict sense, it is necessary for the
matter field to carry an independently conserved charge, which
does not occurs in our case.  

Other point of interest consists in looking for time-dependent
solutions.  The existence of these solutions could show that a
black hole can be regarded as the result of collapsed matter
fields \cite{MannRoss,Husain1,Virbhadra,Husain2,cz}. However,
for the system of massless conformal scalar matter field coupled
to gravity, assuming a stationary spherically symmetric
geometry, ({\em i.e.}, $F=F(r,t)$ and $\Psi=\Psi(r,t)$), gives
rise the same {\em static} solution (Birkhoff's theorem). 

A related question is whether the static solution presented here
is stable under linear perturbations. The question can be
addressed for the case of circularly symmetric perturbations and
will be discussed elsewhere \cite{M-Z}. 

We note that the entropy differs by a factor of $\frac{2}{3}$
from the ``area law" $\frac{\pi}{2} r_+$. This deviation from
the area law was also found in other systems of matter fields
coupled to gravity \cite{CreightonMann}. In \cite{Visser,Englert} this
deviation is also shown to arise in ``dirty" black hole and in
systems of black holes coupled to strings.

Note added: A family of solutions of scalar-tensor fields
coupled to gravity in 2+1 dimensions was recently reported 
\cite{Chan}. It seems that the solution presented here might be
obtained as a special --particularly simple-- case among many
others, but this is not completely clear to these authors at the
moment.

\acknowledgments
Useful discussions with M. Ba\~{n}ados, C. Teitelboim
 and R. Troncoso are
gratefully acknowledged.  This work was supported in part by
Grants Nos.  2940012/94, 4950004/95 1940203 and 1960229 of FONDECYT
(Chile) and Grant No. 27-953/ZI (USACH). The institutional support of
a group of Chilean private companies (EMPRESAS CMPC, CGE, COPEC, MINERA 
ESCONDIDA, NOVAGAS Transportadores de Chile, BUSINESS DESIGN ASS., XEROX
 Chile) is also recognized.



\begin{references}
\bibitem{footnote} In these coordinates, the metric reads
$ds^2=- Fdv^2 + 2dvdr + r^2d\theta^2$. The horizon at
$r=2B\equiv r_+$is generated by the geodesics $ r(\lambda)=r_+
\,\,\, \frac{d\theta}{d \lambda}=0 \,$.This surface is a
marginally trapped since any of these geodesics satisfies $
\frac{d v}{d \lambda}\frac{d r}{d \lambda}=-\frac{r_+^2}{2} 
\left(\frac{d\theta}{d \lambda}\right)^2 \le 0$, so that $r$
decreases or remains constant with increasing $v$.
\end{references}
\end{document}